%% file: arXiv_paper.tex
\documentclass[%
 reprint,
superscriptaddress,
nofootinbib,
 amsmath,amssymb,
 aps,
pra,
]{revtex4-2}

\usepackage{graphicx}
\usepackage{dcolumn}
\usepackage{bm}
\usepackage{hyperref}
\usepackage[mathlines]{lineno}
\usepackage{float}
\usepackage{array}
\usepackage{amsmath,amssymb}
\usepackage{soul}
\usepackage{color}
\usepackage{physics}
\usepackage{booktabs}
\usepackage{siunitx}
\usepackage[normalem]{ulem}
\usepackage{adjustbox}
\usepackage{textcomp}
\usepackage{enumitem}
\usepackage{tabularx}
\usepackage{bbold}
\usepackage{subcaption}
\usepackage{pgfplots}
\usepackage{amsfonts}
\usepackage{braket}
\usepackage{caption}
\usepackage[verbose=true]{epstopdf}
\usepackage[percent]{overpic}
\usepackage{tikz}
\usetikzlibrary{shapes.geometric, arrows.meta, positioning, calc, decorations.pathmorphing, backgrounds, patterns, fit}

\definecolor{amethyst}{rgb}{0.6, 0.4, 0.8}
\definecolor{blue-violet}{rgb}{0.54, 0.17, 0.89}
\hypersetup{
    colorlinks=true,
    urlcolor=black,
    citecolor=black,
    linkcolor=black,
    bookmarks=true,
    bookmarksopen=true,
}

\begin{document}

\title{Identical-Particle Symmetry-Enabled Complete Coherent Control of Ultracold Atomic and Molecular Collisions}

\author{Jing-Chen Zhang}
 \email{jczhang@terpmail.umd.edu}
\affiliation{Institute for Physical Science and Technology, University of Maryland, College Park, Maryland 20742, USA}
\affiliation{Department of Chemistry and Chemical Biology, University of Maryland, College Park, Maryland 20742, USA}
\affiliation{Joint Quantum Institute, University of Maryland and National Institute of Standards and Technology (NIST), College Park, Maryland 20742, USA}
\affiliation{Joint Center for Quantum Information and Computer Science, University of Maryland, College Park,
Maryland 20742, USA}

\author{Adrien Devolder}
\email{adrien.devolder@utoronto.ca}
\affiliation{Chemical Physics Theory Group, Department of Chemistry, and Center for Quantum Information and Quantum Control, University of Toronto, Toronto, Ontario M5S 3H6, Canada}

\author{Timur V. Tscherbul}
\email{ttscherbul@unr.edu}
\affiliation{Department of Physics, University of Nevada, Reno, Nevada 89557, USA}

\author{Paul Brumer}
\email{paul.brumer@utoronto.ca}
\affiliation{Chemical Physics Theory Group, Department of Chemistry, and Center for Quantum Information and Quantum Control, University of Toronto, Toronto, Ontario M5S 3H6, Canada}

\author{Yu Liu}
 \email{yuliu@umd.edu}
\affiliation{Institute for Physical Science and Technology, University of Maryland, College Park, Maryland 20742, USA}
\affiliation{Department of Chemistry and Chemical Biology, University of Maryland, College Park, Maryland 20742, USA}
\affiliation{Joint Quantum Institute, University of Maryland and National Institute of Standards and Technology (NIST), College Park, Maryland 20742, USA}

	\date{\today}

\begin{abstract}
We show that exchange symmetry in collisions of identical particles enables symmetry-protected coherent control of the total scattering cross section. For identical fermions, antisymmetrization enforces a common control phase among contributing channels within a given parity sector, yielding maximal control visibility. For identical bosons, phase-locking persists but with reduced visibility due to additional exchange (satellite) contributions. Collisions of distinguishable particles lack this symmetry-imposed phase-locking, leading to lower controllability and visibility. We elucidate these principles through coupled-channel quantum-scattering calculations for lithium–lithium collisions, comparing the $^{6}\mathrm{Li}{-}^{6}\mathrm{Li}$ (identical fermions), $^{7}\mathrm{Li}{-}^{7}\mathrm{Li}$ (identical bosons), and $^{6}\mathrm{Li}{-}^{7}\mathrm{Li}$ (distinguishable) systems. Furthermore, in the identical-particle cases, symmetry-enforced phase-locking enables full control over the parity of the final state even beyond the ultracold regime. This mechanism is broadly applicable to identical-particle collisions, including homonuclear molecules for which established approaches—DC electric fields or microwave shielding—are ineffective or unavailable.
\end{abstract}

	\maketitle
\section{Introduction}

\begin{figure*}[tbp]
	\centering
	\includegraphics[width=0.85\textwidth]{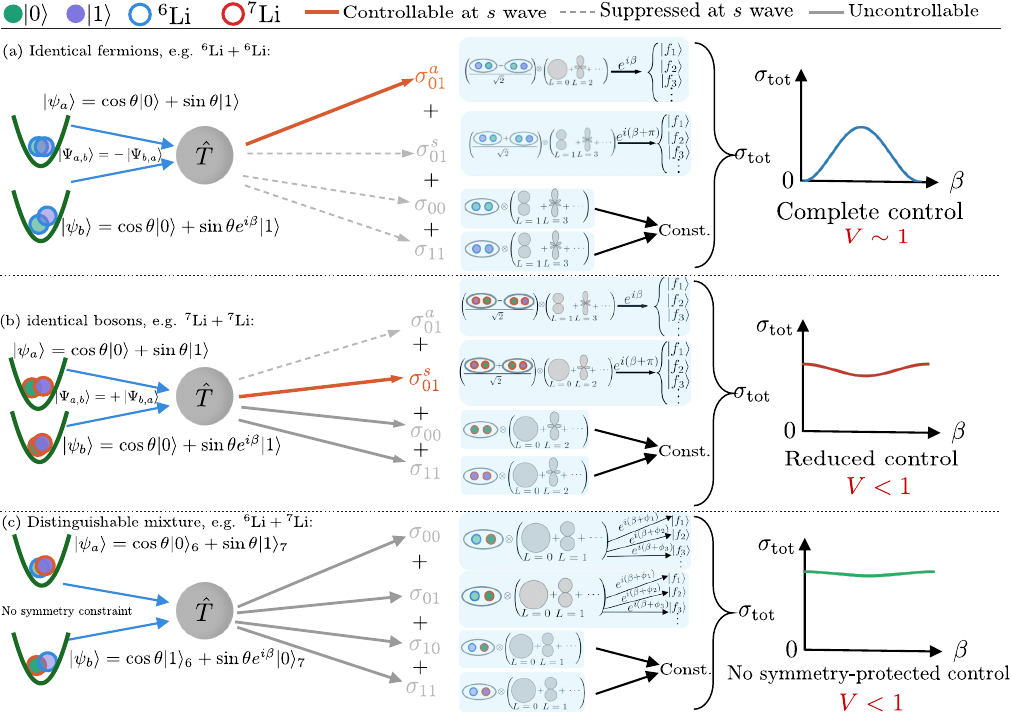}
\caption{Exchange-symmetry–enabled coherent control of the total cross section
$\sigma_{\rm tot}$ for (a) identical fermions ($^{6}$Li$+^{6}$Li), (b) identical
bosons ($^{7}$Li$+^{7}$Li), and (c) a distinguishable mixture
($^{6}$Li$+^{7}$Li). 
\emph{Left}: the atoms are prepared in the superpositions $\ket{\psi_a}$,
$\ket{\psi_b}$ of Eqs.~\eqref{eq:sup_first}--\eqref{eq:sup_second} and
collide via via the T-matrix $\hat T$. \emph{Middle}: each two-particle internal state scatters through its
symmetry-allowed partial waves $L$ into the final channels
$\{\ket{f_i}\}$; summing the state-to-state cross sections over these channels
gives that sector's contribution to $\sigma_{\rm tot}$. \emph{Right}: the exchange-symmetry constraints on the available channels set
the resulting $\sigma_{\rm tot}(\beta)$ and control visibility $V$ for each case.
}
	\label{fig_large}
	\end{figure*}

Over the past few decades, advances in quantum technology have enabled precise preparation, control, and detection of ultracold atoms and molecules \cite{Gross2017Science,Cornish2024,Karman2024NaturePhysics,Carr2009NewJ.Phys.,Bohn2017Science}. External fields, especially magnetic Feshbach resonances, routinely tune atomic scattering with high precision \cite{Chin2010Rev.Mod.Phys.}. While such resonances have also been observed in ultracold molecule-molecule collisions \cite{Park2023}, comparable control for these collisions remains challenging. Nevertheless, state-resolved preparation and detection are increasingly viable, bringing molecular collisions to the forefront of ultracold chemistry and quantum science \cite{Liu2022,Tscherbul2015Phys.Rev.Lett.,Quemener2012ChemicalReviews}.

A central challenge is controlling ultracold bimolecular chemical reactions such as $\mathrm{KRb} + \mathrm{KRb} \rightarrow \mathrm{K_2} + \mathrm{Rb_2}$ \cite{Ni2010Nature,Hu2019Science}, whose many internal and relative-motion degrees of freedom make full-dimensional quantum scattering intractable \cite{Krems2008Phys.Chem.Chem.Phys.}. At short range, dense resonances and many exit channels mean that controlling the \emph{total} cross section demands phase coherence across different scattering channels. Superposition-based coherent control \cite{Book_Brumer,Gong2003} can steer particular state-to-state channels, but the total cross section is usually insensitive: non-interfering ``satellite'' terms from the superposition components \cite{Book_Brumer} and short-range internal-state and partial-wave scrambling wash out interference when summing over final states \cite{Devolder2023}. Entanglement can suppress the satellite background \cite{Devolder2025}, but phase scrambling across final channels remains the main obstacle.

Existing strategies regain control by restricting interference pathways. Symmetry-based methods exploit selection rules (e.g., nuclear-spin sectors), often assuming short-range nuclear-spin conservation \cite{Luke2024,Liu2024Science}, or use time-reversal-invariant eigenstates \cite{Devolder_timerev}. For polar molecules, microwave dressing generates repulsive adiabatic potentials that suppress access to the short-range region and reduce inelastic and reactive losses \cite{Karman2018,Lassabliere2018,Anderegg2021}. Double microwave shielding further enhances this suppression and has recently enabled molecular Bose--Einstein condensation \cite{Bigagli2024,Karman2025,Shi2025}. However, nuclear-spin conservation is not guaranteed in complex-forming collisions \cite{Liu2025NatureChemistry}, and microwave shielding requires a permanent electric dipole, excluding homonuclear molecules.

Thus, a general strategy for controlling the total cross section in collisions of identical ultracold particles remains missing. This is particularly appealing because the total cross section is experimentally accessible and provides a natural observable for the first demonstrations of coherent control in bimolecular collisions \cite{Wurtz_2010,van_der_Poel_2018,Devolder2022}. The coherent control of the total cross section is described by the coherent multichannel optical theorem \cite{Devolder2022}, which specifies conditions for full coherent control in terms of the imaginary parts of forward-scattering amplitudes. However, it does not reveal the underlying physical mechanisms that allow these conditions to be met.

Here we show that identical-particle exchange symmetry provides such a mechanism by fixing the relative phase between exchange-related scattering pathways to 0 or $\pi$ within each exchange-parity sector. This synchronizes the control phase across exit channels, allowing the quantum interference underlying coherent control \cite{Book_Brumer}  to survive summation over final states and enabling control of the total cross section despite complex short-range dynamics (Fig.~\ref{fig_large}). We find maximal visibility for identical fermions, reduced visibility for identical bosons due to residual satellite contributions, and substantially reduced for distinguishable particles. For ultracold $^6$Li+$^6$Li, $^7$Li+$^7$Li, and $^6$Li+$^7$Li collisions, tailored internal superpositions control exchange-parity populations and thus scattering outcomes. This is particularly promising for nonpolar or weakly polar molecules such as $^6$Li$^7$Li+$^6$Li$^7$Li or Sr$_2$+Sr$_2$, where dipole-enabled shielding is unavailable.
    
\section{Limitations of coherent control for distinguishable particles}
Scrambling arising from conflicts between individual control pathways is a fundamental limitation of coherent control in collisions \cite{Devolder2023,Devolder2023TheJournalofPhysicalChemistryLetters}. To clarify its origin and how exchange symmetry can mitigate it, we first consider collisions of non-identical particles, where molecules A and B are independently prepared in coherent superpositions of two internal eigenstates, $\ket{0}$ and $\ket{1}$:
\begin{equation}
	\ket{\psi_a}=\cos \theta \ket{0}_A + \sin \theta  \ket{1}_A
    \label{eq:sup_first}
\end{equation}
\begin{equation}
	\ket{\psi_b}=\cos \theta \ket{0}_B + \sin \theta e^{i\beta}  \ket{1}_B
    \label{eq:sup_second}
\end{equation}
with a controllable relative phase $\beta$ introduced between the two molecules. Quantum interference requires degenerate internal states; otherwise, $\ket{0}_A\ket{1}_B$ and $\ket{1}_A\ket{0}_B$ are energetically distinguishable and do not interfere \cite{Brumer2000}. In practice, this is often achieved using superpositions of states with the same internal angular momentum but different projections, i.e., $m$-superpositions \cite{Brumer1999}.

The corresponding total internal state of the two-molecule system can be expressed as the tensor product of the individual states: $\ket{\Psi_{\mathrm{ini}}} =\cos^2\theta \ket{0}_A\ket{0}_B + \cos\theta \sin\theta e^{i\beta} \ket{0}_A\ket{1}_B+\cos\theta \sin\theta \ket{1}_A\ket{0}_B+\sin^2\theta e^{i\beta} \ket{1}_A\ket{1}_B$.
Alternatively, one can prepare entangled superpositions: $\ket{\Psi_{\mathrm{ini}}}=\cos\theta \ket{0}_A\ket{1}_B+  \sin \theta e^{i\beta} \ket{1}_A\ket{0}_B$, which avoid population in the uncontrolled satellite states $\ket{0}_A\ket{0}_B$ and $\ket{1}_A\ket{1}_B$ and are expected to offer enhanced controllability \cite{Gong2003,Devolder2025}. Although such entangled molecular states have recently been realized \cite{Bao_2023,Picard_2024,Ruttley_2025}, they remain experimentally challenging. We therefore focus on product states, with entangled-state results given in Appendix~\ref{sec:app_initial_states}. 

The total scattering cross section can be written as $\sigma=\boldsymbol{a}^\dagger \boldsymbol{C} \boldsymbol{a}$ \cite{Frishman1999,Devolder2024TheJournalofChemicalPhysics}, where $\boldsymbol{a}=(\cos^2\theta, \cos\theta \sin\theta e^{i\beta}, \cos\theta \sin\theta, \sin^2\theta e^{i\beta})$. The coherent control scattering (CCS) matrix $\boldsymbol{C}$ is defined as
\begin{equation}
\boldsymbol{C}_{ij}=\frac{\pi}{k^2}\sum_f \sum_{\ell,m} \sum_{\ell',m'} T_{i,\ell,m\rightarrow f,\ell',m'}T_{j,\ell,m\rightarrow f,\ell',m'}^* , 
\label{eq:CCS}
\end{equation} 
in terms of $T_{i,\ell,m\rightarrow f,\ell',m'}$, the $T$-matrix elements connecting the initial internal state $\ket{i}$ in the initial partial wave ($\ell,m$) to the final state $\ket{f}$ in the final partial wave ($\ell',m'$). Here, $m$ and $m'$ are the projections of the partial-wave states $\ell$ and $\ell'$ on the quantization axis, and $k$ is the relative incident momentum. $\ket{i}\equiv\ket{i_A}_A\ket{i_B}_B$, where $\ket{i_A}_A$ and $\ket{i_B}_B$ are the initial internal states of molecules A and B, respectively. $\ket{f}\equiv\ket{f_A}_A\ket{f_B}_B$, where $\ket{f_A}_A$ and $\ket{f_B}_B$ are the post-collision internal states of molecules A and B, respectively. The sum over final states includes the elastic channel as well as all inelastic transitions. Introductions to coherent control and the CCS matrix are given in Appendices~\ref{sec:app_general_intro} and \ref{sec:app_CCS}.

We note that nonvanishing interference requires equal total projection of the internal angular momentum and energy \cite{Omiste2018}, so only the states $\ket{0}_A\ket{1}_B$ and $\ket{1}_A \ket{0}_B$ interfere. The resulting total cross section is:
\begin{equation}
\begin{split}
\sigma^{\mathrm{non-ident}}(\theta, \beta)= &\cos^4\theta \sigma_{0,0}+\cos^2\theta \sin^2\theta \sigma_{0,1}\\
+&\cos^2\theta \sin^2\theta \sigma_{1,0} + \sin^4\theta \sigma_{1,1}\\
+& 2\sin^2\theta \cos^2\theta  \sigma_{\mathrm{int}} (\beta)
\end{split}
\label{eq:cont_total_non_ident}
\end{equation}
where each $\sigma$ is the diagonal element of the CCS matrix, e.g., $\sigma_{0,0}=\boldsymbol{C}_{00,00}=\frac{\pi}{k^2}\sum_{f=1}^{N_f} \sum_{\ell,m} \sum_{\ell',m'} |T_{00,\ell,m\rightarrow f,\ell',m'}|^2$. $\sigma_{\mathrm{int}}(\beta)$ is the total interference term between $\ket{01}$ and $\ket{10}$ and is defined as:
\begin{equation}
\begin{split}
\sigma_{\mathrm{int}} (\beta)=\sum_f \sum_{\ell,m,\ell',m'} & |\sigma_{\mathrm{int},\ell,m \rightarrow f,\ell',m'}| \\ &\times \cos(\beta+\Phi_{f,\ell,m,\ell',m'})
\end{split}
\label{eq: interference_term_non_ident}
\end{equation}
 Here, $\sigma_{\mathrm{int},\ell,m \rightarrow f,\ell',m'}=\frac{\pi}{k^2} T_{01,\ell,m\rightarrow f,\ell',m'}T_{10,\ell,m\rightarrow f,\ell',m'}^* $ is the individual interference term for the specific final internal states $\ket{f}$ and partial waves $(\ell,m,\ell',m')$. $\Phi_{f,\ell,m,\ell',m'}$ is the phase of $\sigma_{\mathrm{int},\ell,m \rightarrow f,\ell',m'}$  and corresponds to the relative phase between $T_{01,\ell,m\rightarrow f,\ell',m'}$ and $T_{10,\ell,m\rightarrow f,\ell',m'}$ .

Summing the contributions in Eq.~\eqref{eq: interference_term_non_ident} reduces controllability due to two scrambling mechanisms: phase averaging and path distinguishability. If the relative phases $\Phi_{f,\ell,m,\ell',m'}$ are random, interference terms cancel upon summation, leading to dephasing and loss of control \cite{Devolder2023TheJournalofPhysicalChemistryLetters,Devolder2023}. Moreover, maximal interference requires indistinguishable pathways, but the ratio $\frac{|T_{01,\ell,m\rightarrow f,\ell',m'}|^2}{|T_{10,\ell,m\rightarrow f,\ell',m'}|^2}$ depends on the final state and partial waves, so no choice of initial populations can enforce indistinguishability for all channels simultaneously.

Scrambling from summation over initial and final partial waves is termed partial-wave scrambling, while that from summing over final internal states is internal-state scrambling. At ultracold temperatures, the limited number of contributing partial waves reduces partial-wave scrambling \cite{Devolder2021Phys.Rev.Lett.}, but internal-state scrambling remains and continues to limit control. Thus, coherent control is not forbidden for distinguishable particles, but is not protected against scrambling when the total cross section is formed. The scrambling reduces the magnitude of the total interference term $\sigma_{\mathrm{int}}$ relative to the geometric mean of the cross sections, $\sqrt{\sigma_{0,1}\sigma_{1,0}}$, and consequently reduces the Cauchy–Schwarz-related quantity $\frac{|\sigma_{\mathrm{int}}|}{\sqrt{\sigma_{0,1}\sigma_{1,0}}}$, characterizing the extent of control.

\section{Coherent control of identical particles}
These limitations from partial-wave and internal-state scrambling can be removed in ultracold collisions of identical particles via exchange symmetry, as shown below. Exchange symmetry locks the two-particle internal-state symmetry to the partial-wave parity, allowing only symmetry-allowed partial waves within each sector and constraining exchange-related scattering amplitudes to differ only by relative phases of $0$ or $\pi$. As a result, the coherent-control interference term survives summation over final states and same-parity partial waves, enabling control of the total cross section.

Unlike the non-identical case, the single-particle states $\ket{0}$ and $\ket{1}$ in Eqs.~\eqref{eq:sup_first} and \eqref{eq:sup_second} need not be degenerate \cite{Brumer2000}, relaxing constraints on preparing $m$-superpositions. However, both particles must be prepared in the same pair of basis states; otherwise exchange symmetry is broken and channel-dependent relative phases reappear (see Appendix~\ref{sec:app_different states}). For identical particles, it is convenient to rewrite the two-molecule state in symmetric and antisymmetric combinations of $\ket{0}\ket{1}$ and $\ket{1}\ket{0}$: $\ket{0,1}_s=\frac{1}{\sqrt{2}}\left(\ket{0}\ket{1}+\ket{1}\ket{0}\right)$ and $\ket{0,1}_a=\frac{1}{\sqrt{2}}\left(\ket{0}\ket{1}-\ket{1}\ket{0}\right)$. Using these definitions, the initial two-molecule state can be rewritten as $\ket{\Psi_{\mathrm{ini}}} =\cos^2\theta \ket{0}\ket{0}+\frac{\sin\theta \cos\theta}{\sqrt{2}} (1+e^{i\beta})\ket{0,1}_s+\frac{\sin\theta \cos\theta}{\sqrt{2}} (e^{i\beta}-1)\ket{0,1}_a+\sin^2\theta e^{i\beta} \ket{1}\ket{1}$. The states $\ket{0}\ket{0},\ket{0,1}_s$ and $\ket{1}\ket{1}$ are symmetric under particle exchange. Consequently, fermionic molecules in these states scatter via odd partial waves, while bosonic molecules scatter via even partial waves. In contrast, the antisymmetric state $\ket{0,1}_a$ scatters via even partial waves for fermions and odd partial waves for bosons.

Interference in the symmetrized basis is determined by the corresponding CCS matrix. It can occur only between states with identical internal energies, restricting it to $\ket{0,1}_s$ and $\ket{0,1}_a$. However, the CCS matrix element between these two states vanishes: a nonzero contribution requires the corresponding $T$-matrix elements to be nonzero for identical initial and final partial waves. Because symmetric and antisymmetric states correspond to opposite partial-wave parities, no interference survives in the basis $(\ket{0,0}, \ket{0,1}_s, \ket{0,1}_a, \ket{1,1})$. The absence of off-diagonal interference in the symmetry-adapted basis does not imply an absence of coherent interference. In the unsymmetrized basis, the same physics appears as maximal interference between exchange-related pathways within each parity sector (Appendix~\ref{sec:app_unsymmetrized_basis}); the symmetry-adapted representation converts this interference into phase-dependent population redistribution between orthogonal exchange sectors \cite{Devolder2024TheJournalofChemicalPhysics}.

Since the CCS matrix is diagonal, the total cross section from the product-state superposition becomes:
\begin{equation}
\begin{split}
\sigma^{\mathrm{ident}}(\theta,\beta)=
&\cos^4\theta \sigma_{0,0}+\cos^2\theta \sin^2\theta \sigma^s_{0,1}\\
+&\cos^2\theta \sin^2\theta \sigma^a_{0,1} + \sin^4\theta \sigma_{1,1}\\
+&2\cos^2\theta \sin^2\theta \sigma_{\mathrm{int}}(\beta)\\
\end{split}
\label{eq:cont_total_ident}
\end{equation}
where the total interference term $\sigma_{\mathrm{int}}(\beta)$ takes the simple form:
\begin{equation}
\sigma_{\mathrm{int}}(\beta)= \frac{\sigma^s_{0,1}}{2} \cos(\beta)+\frac{\sigma^a_{0,1}}{2} \cos(\beta+\pi)
\label{eq: interference_term_ident}
\end{equation}

Coherent control of the total cross section involves four contributions: $\sigma_{0,0}$, $\sigma^s_{0,1}$, $\sigma^a_{0,1}$, and $\sigma_{1,1}$. The two satellite terms are independent of $\beta$, while the symmetric and antisymmetric terms vary oppositely with $\beta$ (maximized at $\beta=0$ or $\pi$, respectively). Coherent control thus arises from the competition between the symmetric and antisymmetric contributions.

In the ultracold regime, scattering is dominated by the s-wave. For identical fermions, only the antisymmetric channel contributes in the $s$-wave limit:
\begin{equation}
	\sigma ^{\mathrm{fermion}} (\theta, \beta) \approx \frac{\cos^2\theta \sin^2\theta}{2} |1-e^{i\beta}|^2 \sigma^a_{0,1}
    \label{eq:control_fermion}
\end{equation}
yielding complete control from 0 ($\beta=0$) to $2\cos^2\theta\sin^2\theta\,\sigma^a_{0,1}$ ($\beta=\pi$). The symmetry-enforced phase locking ensures that the magnitude of the interference term equals the incoherent terms, thereby saturating the Cauchy–Schwarz inequality, as discussed further in the unsymmetrized basis in Appendix \ref{sec:app_unsymmetrized_basis}. For identical bosons, the symmetric channel and satellite states contribute via $s$-wave scattering:
\begin{equation}
\begin{split}
\sigma  ^{\mathrm{bosons}} (\theta, \beta) \approx & \cos^4\theta \sigma_{0,0}+\frac{\cos^2\theta \sin^2\theta}{2} |1+e^{i\beta}|^2 \sigma^s_{0,1} \\ +& \sin^4\theta \sigma_{1,1}
\end{split}
    \label{eq:control_boson}
\end{equation}
Control remains synchronized, but extrema are reversed ($\beta=0$ maximum, $\beta=\pi$ minimum) and satellite terms can weaken the control. Complete control can be achieved using entangled superpositions in which satellite contributions are absent, as demonstrated in Appendix \ref{sec:app_initial_states}.

The extent of control is quantified by the visibility~\cite{Devolder2024TheJournalofChemicalPhysics,Devolder2023}:
\begin{equation}
\begin{split}
V(\theta)= &\frac{\sigma_{\mathrm{max}}(\theta,\beta_{\mathrm{max}})-\sigma_{\mathrm{min}}(\theta,\beta_{\mathrm{min}})}{\sigma_{\mathrm{max}}(\theta,\beta_{\max})+\sigma_{\min}(\theta,\beta_{\min})} \\
=&\frac{\left|\frac{\sigma_{01}^s}{\sigma_{01}^a}-1\right|}{\cot^2\theta\frac{\sigma_{00}}{\sigma_{01}^a}+\tan^2\theta \frac{\sigma_{11}}{\sigma_{01}^a}+\frac{\sigma_{01}^s}{\sigma_{01}^a}+1},
\end{split}
\label{eq:visibility}
\end{equation}
which depends on the ratio of symmetric to antisymmetric contributions, $\sigma_{01}^s/\sigma_{01}^a$, and the weights of the satellite terms, $\sigma_{00}/\sigma_{01}^a$ and $\sigma_{11}/\sigma_{01}^a$. Visibility is high when one contribution dominates (e.g., in the ultracold regime) and vanishes when $\sigma_{01}^s \approx \sigma_{01}^a$, reflecting the competition between symmetric and antisymmetric contributions.

Comparing Eqs.~\eqref{eq:cont_total_non_ident}--\eqref{eq: interference_term_non_ident} for non-identical particles with Eqs.~\eqref{eq:cont_total_ident}--\eqref{eq: interference_term_ident} for identical particles shows that exchange symmetry removes key limitations of coherent control. In the non-identical case, relative phase varies across final states and partial waves, leading to partial-wave and internal-state scrambling upon summation. For identical particles, exchange symmetry restricts relative phases to 0 or $\pi$ set by scattering parity. In the ultracold regime, a single parity typically dominates, suppressing scrambling and enforcing pathway indistinguishability.

\section{Symmetry-synchronized coherent control of lithium isotopologues in the ultracold regime}

\begin{figure*}[t]
	\centering
	\includegraphics[width= 0.85 \textwidth]{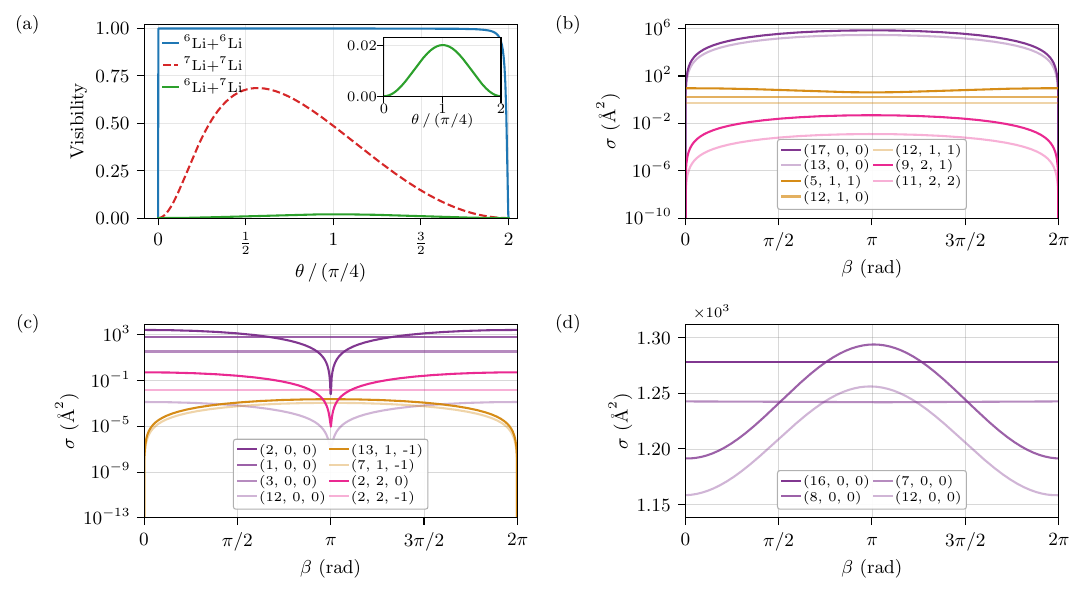}
\caption{(a) Visibility $V$ of the total cross-section control versus $\theta$ for $^{6}$Li+$^{6}$Li, $^{7}$Li+$^{7}$Li, and $^{6}$Li+$^{7}$Li at $E=20\,\mu\mathrm{K}$; the inset zooms in for $^{6}$Li+$^{7}$Li.
(b)--(d) Dominant partial-wave-resolved state-to-state cross sections for (b) $^{6}$Li+$^{6}$Li, (c) $^{7}$Li+$^{7}$Li, and (d) $^{6}$Li+$^{7}$Li.
Labels denote exit-channel quantum numbers $(\alpha_{\rm out},L_{\rm out},m_{L,\rm out})$; each $\alpha_{\rm out}$ labels one specific combination of internal quantum numbers used in \textsc{MOLSCAT}~\cite{Hutson2019ComputerPhysicsCommunications}, e.g., $(f_A,m_{f_A},f_B,m_{f_B})$ for two-atom collisions; colors indicate partial waves and opacity indicates hyperfine exit channels. The flat traces of the state-to-state cross sections are due to satellite terms.}
	\label{fig:coherent_control}
	\end{figure*}

To illustrate symmetry-enforced synchronization of coherent control, we perform zero-field coupled-channel calculations for $^{6}$Li+$^{6}$Li (identical fermions), $^{7}$Li+$^{7}$Li (identical bosons), and $^{6}$Li+$^{7}$Li (distinguishable), comparing identical-particle collisions to a heteronuclear case without symmetry-locked phases. We model ground-state alkali-atom collisions as in Ref.~\cite{Zhang2026Feshbach} and compute multichannel $T$-matrix elements using \textsc{MOLSCAT} \cite{Hutson2019ComputerPhysicsCommunications} at $E=20~\mu\mathrm{K}$ unless stated otherwise; further details are in Appendix~\ref{sec:app_basis_channels}. From these, we construct CCS matrix elements (Eq.~\eqref{eq:CCS}) in unsymmetrized and symmetrized bases for non-identical and identical cases, respectively, and evaluate $\sigma=\boldsymbol{a}^\dagger \boldsymbol{C} \boldsymbol{a}$.

The Li atoms are prepared in the coherent superposition given in Eqs.~\eqref{eq:sup_first} and \eqref{eq:sup_second}. To illustrate the control, we use the following representative internal states: \(\ket{0}\equiv\ket{f_{Li}=1/2,m_{f_{Li}}=1/2}_{{}^{6}\mathrm{Li}}\) and
\(\ket{1}\equiv\ket{f_{Li}=3/2,m_{f_{Li}}=1/2}_{{}^{6}\mathrm{Li}}\) for homonuclear $^{6}$Li+$^{6}$Li collisions, \(\ket{0}\equiv\ket{f_{Li}=1,m_{f_{Li}}=-1}_{{}^{7}\mathrm{Li}}\) and
\(\ket{1}\equiv\ket{f_{Li}=2,m_{f_{Li}}=-2}_{{}^{7}\mathrm{Li}}\) for homonuclear
$^{7}$Li+$^{7}$Li collisions. For heteronuclear collisions, we choose \(\ket{0}_6\equiv\ket{f_{Li}=1/2,m_{f_{Li}}=1/2}_{{}^{6}\mathrm{Li}}\),
\(\ket{1}_6\equiv\ket{f_{Li}=1/2,m_{f_{Li}}=-1/2}_{{}^{6}\mathrm{Li}}\),
\(\ket{0}_7\equiv\ket{f_{Li}=1,m_{f_{Li}}=1}_{{}^{7}\mathrm{Li}}\), and
\(\ket{1}_7\equiv\ket{f_{Li}=1,m_{f_{Li}}=0}_{{}^{7}\mathrm{Li}}\). Here, $f_{Li}$ and $m_{f_{Li}}$ are the quantum numbers defining the hyperfine states of Li atoms. The final channels populated by collisions between atoms prepared in these superpositions correspond to elastic scattering, spin-exchange processes, or dipolar relaxation into lower hyperfine states. A practical route to prepare these superpositions uses two initially separated atomic ensembles (tweezers or optical traps) that are independently prepared and then merged for collisions. Similar schemes have been proposed for ultracold-molecule interferometry, where site-selective AC Stark shifts and phase-coherent microwave pulses imprint a controllable relative phase before merging \cite{Luke2024}. For lithium atoms, rf/microwave fields and differential light shifts can set the control phase $\beta$ between entrance pathways. The effectiveness of this control is sensitive to any dephasing that may occur between state preparation and collision.

\begin{figure*}[tbp]
\centering
\includegraphics[width=0.85 \textwidth]{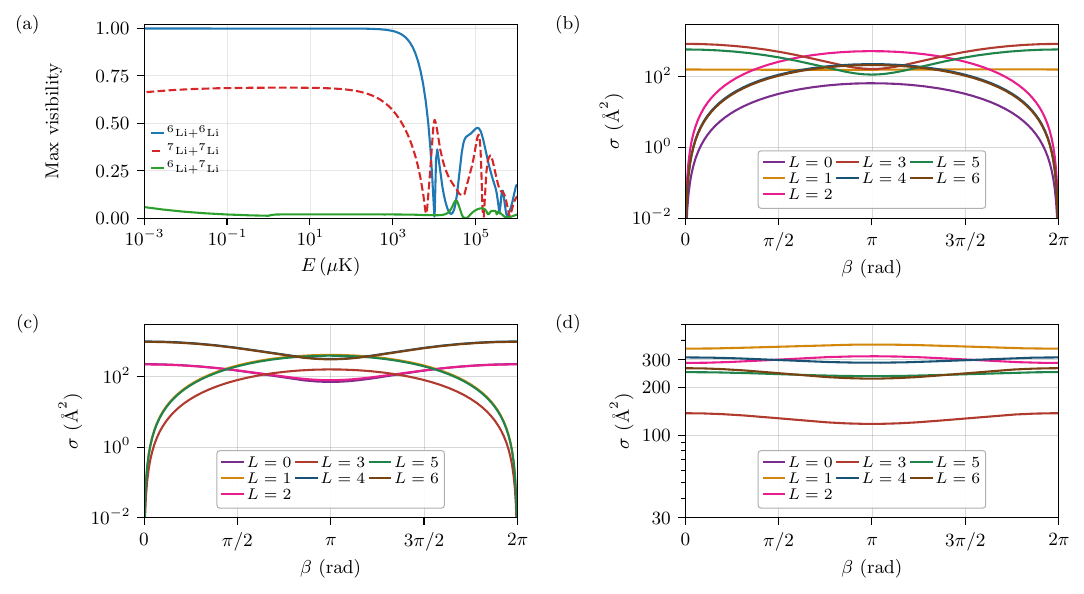}
\caption{(a) Maximum visibility $V$ of the total cross-section control versus collision energy $E$ for $^{6}$Li+$^{6}$Li, $^{7}$Li+$^{7}$Li, and $^{6}$Li+$^{7}$Li.
(b)--(d) Total cross sections at different partial waves at $E=1\,\mathrm{K}$ for
(b) $^{6}$Li+$^{6}$Li, (c) $^{7}$Li+$^{7}$Li, and (d) $^{6}$Li+$^{7}$Li.}
\label{fig:visibility_and_partials}
\end{figure*}

The visibilities for fermionic, bosonic, and non-identical cases are compared in Fig.~\ref{fig:coherent_control}(a). For fermionic $^{6}$Li+$^{6}$Li collisions, control is near-perfect ($V \simeq 1$), consistent with Eq.~\eqref{eq:control_fermion}: symmetry suppresses satellite and symmetric contributions, leaving the antisymmetric channel dominant and producing synchronized control with a minimum at $\beta=0$ [Fig.~\ref{fig:coherent_control}(b)].
For bosonic $^{7}$Li+$^{7}$Li, control is reduced but still significant ($V=0.7$) due to satellite contributions, though symmetry still synchronizes the main channel with a shifted minimum near $\pi$ [Fig.~\ref{fig:coherent_control}(c)]. Entangled initial states can suppress satellite effects and restore near-unity visibility (see Appendix~\ref{sec:app_initial_states}). In contrast, the distinguishable $^{6}$Li+$^{7}$Li mixture shows very weak control ($V\simeq0.01$) because pathway indistinguishability cannot be enforced across all channels (as shown in Fig.~\ref{fig:6Li7Li_distinguishability} in Appendix~\ref{sec:app_initial_states}). 

\section{Beyond the ultracold regime}

As collision energy increases, more partial waves become accessible, but exchange symmetry in identical particles still enforces locked relative phases within each parity sector. As a result, no partial-wave or hyperfine scrambling occurs within a given parity sector. This is illustrated in Fig.~\ref{fig:visibility_and_partials}(b) and (c) at $1\,\mathrm{K}$, where contributions are summed over hyperfine states for fixed partial waves: even and odd channels each show synchronized control. Tuning $\beta$ allows switching between regimes dominated by even or odd partial waves. This parity-selective control persists across the calculated collision-energy range and is a hallmark of coherent control in identical-particle scattering.

The parity switch is possible because contributions from opposite-parity channels are out of phase. When summed, however, they partially cancel, reducing the visibility of the total cross section [Fig.~\ref{fig:visibility_and_partials}(a)]. The visibility vanishes when the contributions from the two parities become equal and is directly related to the ratio $\sigma_{01}^s/\sigma_{01}^a$ [Eq.~\eqref{eq:visibility}] (see further analysis in Appendix~\ref{sec:app_energy_sync}). This condition is reached beyond the $p$-wave centrifugal barrier (approximately $10\,\mathrm{mK}$), above which $s$- and $p$-wave contributions become comparable.

In contrast, heteronuclear cases exhibit both hyperfine and partial-wave 
scrambling. Hyperfine scrambling reduces control when summing over 
hyperfine states at fixed partial wave, as shown in Fig.~\ref{fig:totaltheta_entangled}(d) of Appendix~\ref{sec:app_initial_states}. Partial-wave scrambling, shown in 
Fig.~\ref{fig:visibility_and_partials}(d), arises within each parity sector 
from the lack of phase-locking: channels such as $L=1$ and $L=3$ (and $L=2$ 
and $L=6$) are out of phase. Consequently, tuning $\beta$ controls neither the 
parity nor the total cross section in the heteronuclear case.

\section{Conclusions}
We have shown that exchange symmetry of identical collision partners synchronizes the coherent control of state-to-state cross sections of different exit channels with the same partial-wave parity. This property enables strong theoretical control of bimolecular collisions. In the ultracold regime, this enables strong control of the total cross section, reaching complete control when a single symmetry sector dominates and uncontrolled satellite contributions are absent or suppressed. Beyond the ultracold regime, although control over the total cross sections is substantially reduced, the distinct phase locking within each parity sector enables controlled manipulation of the final-state parity. By overcoming phase-scrambling limitations of previous approaches, this work demonstrates coherent control of a total cross section, with control over the parity of the exit channels remaining robust at higher collision energies.

Symmetry-synchronized control in Li+Li collisions provides a starting point for extending coherent-control strategies to molecular systems. Our protocol is particularly useful for homonuclear systems, which lack permanent dipole moments and therefore cannot be controlled using external electric fields, and for which short-range dynamics are complex and microwave shielding is ineffective. The protocol applies broadly to identical particles and is not intrinsically degraded by an increasing number of final internal states, opening new avenues for controlling collisions such as Li$_2$+Li$_2$ and Sr$_2$+Sr$_2$.

\section*{Acknowledgments} 
We thank the participants of the 2025 Gordon Research Conference on Quantum Control for valuable discussions. This work was supported by the U.S. Air Force Office of Scientific Research (AFOSR) under Contract No. FA9550-22-1-0361. Jing-Chen Zhang and Yu Liu are supported by the startup fund of the Department of Chemistry at the University of Maryland, College Park. Jing-Chen Zhang also thanks Kang-Kuen Ni and Kaden Hazzard for valuable discussions.

\appendix

\section{Details of coupled-channel calculations}
\label{sec:app_basis_channels}

\subsection{Hamiltonian and interaction potentials}
\label{sec:hamiltonian_potentials}

The Hamiltonian describing the relative motion of two Li atoms in their electronic ground state is
\begin{equation}
\label{eq:hamiltonian}
\hat H
=
-\frac{\hbar^{2}}{2\mu}\frac{1}{R}\frac{d^{2}}{dR^{2}}R
+
\frac{\hat L^{2}}{2\mu R^{2}}
+
\hat h_a + \hat h_b
+
\hat V(R),
\end{equation}
where \(\mu\) is the reduced mass, \(R\) is the internuclear separation, and \(\hat L\) is the relative orbital angular momentum operator. The single-atom Hamiltonians \(\hat h_{j}\) for $j\in\{a,b\}$ include hyperfine and Zeeman interactions,
\begin{equation}
\label{eq:singleH}
\hat h_j
=
\zeta_j\,\hat{\boldsymbol{\imath}}_j \cdot \hat{\boldsymbol{s}}_j
+
\left(
g_e \mu_{\mathrm{B}} \hat s_{zj}
+
g_{i,j} \mu_{\mathrm{N}} \hat \imath_{zj}
\right) B,
\end{equation}
where \(\zeta_j\) is the hyperfine constant and \(g_e\) and \(g_{i,j}\) are the electron and nuclear $g$-factors.

The interaction potential \(\hat V(R)\) includes the isotropic adiabatic electronic potentials and the anisotropic magnetic dipole-dipole interaction $\hat V^{\mathrm{d}}(R)$:
\begin{equation}
\label{eq:potdecomp}
\hat V(R) = V_0(R)\,\hat P_0 + V_1(R)\,\hat P_1 + \hat V^{\mathrm{d}}(R) + \sum_{S=0}^{1} V_{\mathrm{shift},S}(R).
\end{equation}
Here, $\hat P_0$ and $\hat P_1$ project onto the electronic singlet ($S=0$) and triplet ($S=1$) subspaces, respectively. The dipole term \(\hat V^{\mathrm{d}}(R)\), dominated by spin-spin coupling, induces weak anisotropic couplings with $\Delta L=\pm 2$ and is given by
\begin{equation}
\hat{V}^{\mathrm{d}}(R)
=
\frac{E_{\mathrm{h}}\alpha^{2}}{(R/a_{0})^{3}}
\left[
\hat{s}_{a}\cdot\hat{s}_{b}
-
3(\hat{s}_{a}\cdot\mathbf{e}_{R})(\hat{s}_{b}\cdot\mathbf{e}_{R})
\right].
\end{equation}

We represent the singlet ($X^1\Sigma^+$) and triplet ($a^3\Sigma^+$) Li\(_2\) potentials using spectroscopically fitted Morse/Long-Range (MLR) potentials~\cite{Gunton2013Phys.Rev.A,Semczuk2013Phys.Rev.A} with correction terms \cite{Zhang2026Feshbach}:
\begin{equation}
\label{eq:MLR}
V_{\mathrm{MLR}}(R) = D_e \left[1 - \frac{u_{\mathrm{LR}}(R)}{u_{\mathrm{LR}}(R_e)}\, e^{-\beta(R)\, y_p(R)}\right]^2,
\end{equation}
where $D_e$ is the dissociation energy, $u_{\mathrm{LR}}$ enforces the correct long-range van der Waals behavior, and the polynomial $\beta(R)$ controls the short-range shape. 

We apply a quadratic correction term to the inner wall ($R < R_{e,S}$) of the adiabatic potentials:
\begin{equation}
\label{eq:shift}
V_{\mathrm{shift},S}(R) = \mathcal{S}_{S}\,(R-R_{e,S})^{2}.
\end{equation}
The shift parameter $\mathcal{S}_{S}$ enables fine tuning of scattering lengths and Feshbach resonance locations. Details of the potential parameters, including the shift terms, can be found in Ref.~\cite{Zhang2026Feshbach}.

We solve the coupled-channel equations using \textsc{MOLSCAT}~\cite{Hutson2019ComputerPhysicsCommunications}. The total Hamiltonian is block-diagonal in sectors labeled by the total angular momentum projection $M_{\rm tot}$ and parity $(-1)^L$. We use a propagator method to solve the scattering wavefunctions in each symmetry sector. The propagation uses the hybrid log-derivative/Airy method from the inner wall to long range. In the short-range region, we use a fixed radial step size $dr=0.002\,a_0$, while the Airy propagator employs its standard adaptive step-size control at long range. We propagate from $R=1.8\,a_0$ to $R=3\times10^{5}\,a_0$, and verify numerical convergence by varying the propagator parameters. We use $L_{\max}=2$ for $0.5\,\mathrm{nK}\leq E\leq1\,\mu\mathrm{K}$ in the ultracold regime and $L_{\max}=6$ for calculations at higher collision energies, $1\,\mu\mathrm{K}\leq E\leq1\,\mathrm{K}$, to ensure an accurate description of shape resonances in Li+Li collisions at high energies.

\subsection{Channel ordering and mapping to hyperfine labels}
\label{sec:app_channel_ordering}

\begin{figure}[tbp]
\centering
\includegraphics[width=\columnwidth]{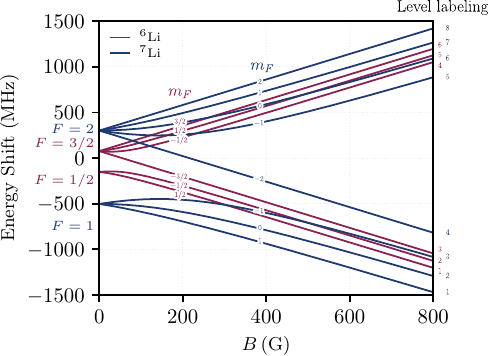}
\caption{Zeeman shifts of the hyperfine levels in the electronic ground state $({}^2S_{1/2})$ of ${ }^6 \mathrm{Li}$ (red) and ${ }^7 \mathrm{Li}$ (blue), labeled by the corresponding quantum numbers $f$ and $m_f$. The zero of energy is defined as the energy in the absence of hyperfine and Zeeman interactions. The numbers on the right label the atomic hyperfine states used to specify each atom in a scattering channel at nonzero $B$.}
\label{fig:channels}
\end{figure}

For convenience, we label the scattering channels by the adiabatic hyperfine states of Li atoms at $B>0$, as shown in Fig.~\ref{fig:channels}. The asymptotic states are ordered by increasing threshold energies at $B=0.1\,\mathrm{G}$ (the magnetic field value used for all scattering calculations throughout the paper, unless stated otherwise). We thus define a two-level qubit manifold $\{\ket{0},\ket{1}\}$ for each collision case ($^{6}$Li+$^{6}$Li, $^{7}$Li+$^{7}$Li, and $^{6}$Li+$^{7}$Li) by mapping the specific states used for coherent control to these adiabatic hyperfine states, as summarized in Table~\ref{tab:twolevel_map}.

Specific hyperfine states were selected to illustrate how control behavior varies across isotopologues. For the homonuclear systems ($^{6}$Li+$^{6}$Li and $^{7}$Li+$^{7}$Li), we chose scattering channels that exhibit strong inelastic relaxation to other adiabatic hyperfine channels; this often complicates magnetic-field-dependent scattering processes and may degrade control by increasing the number of populated channels. Conversely, for the heteronuclear $^{6}$Li+$^{7}$Li system, channels were chosen for their simpler bound-state spectrum in a magnetic field, which facilitates better control. We emphasize, however, that weaker coupling to open channels does not intrinsically guarantee higher visibility in the total scattering cross section. Although we have the freedom to choose other entrance channels, the conclusions regarding identical-particle symmetry presented here are general and independent of the specific entrance channel selected.

\begin{table}[b]
    \centering
    \caption{The two-level manifold $\{\ket{0},\ket{1}\}$ used for coherent control, specified by the approximate single-atom hyperfine quantum numbers $\ket{f,m_f}$ at $B=0.1~\mathrm{G}$; see Fig.~\ref{fig:channels} for the corresponding asymptotic level numbers at different $B$.}
    \begin{tabular}{cccc}
    \hline\hline
    System & Label & ${}^6$Li $(\ket{f,m_f})$ & ${}^7$Li $(\ket{f,m_f})$ \\
    \hline
    ${}^6$Li$+{}^6$Li
      & $\ket{0}$ & $\ket{\tfrac{1}{2},+\tfrac{1}{2}}$ (level $1$) & --- \\
      & $\ket{1}$ & $\ket{\tfrac{3}{2},+\tfrac{1}{2}}$ (level $5$) & --- \\
    ${}^7$Li$+{}^7$Li
      & $\ket{0}$ & --- & $\ket{1,-1}$ (level $3$) \\
      & $\ket{1}$ & --- & $\ket{2,-2}$ (level $4$) \\
    ${}^6$Li$+{}^7$Li
      & $\ket{0}$ & $\ket{\tfrac{1}{2},+\tfrac{1}{2}}$ (level $1$) & $\ket{1,1}$ (level $1$) \\
      & $\ket{1}$ & $\ket{\tfrac{1}{2},-\tfrac{1}{2}}$ (level $2$) & $\ket{1,0}$ (level $2$) \\
    \hline\hline
    \end{tabular}
    \label{tab:twolevel_map}
\end{table}

\begin{figure*}[tbp]
    \centering
    \begin{overpic}[width=0.67\columnwidth]{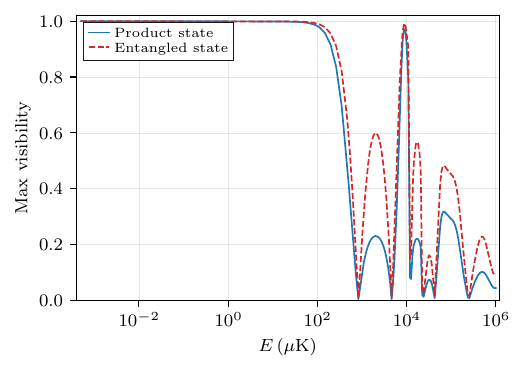}
        \put(0,63){(a)}
    \end{overpic}
    \begin{overpic}[width=0.67\columnwidth]{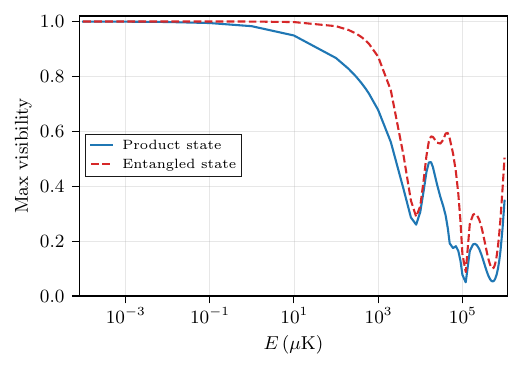}
        \put(0,63){(b)}
    \end{overpic}
    \begin{overpic}[width=0.67\columnwidth]{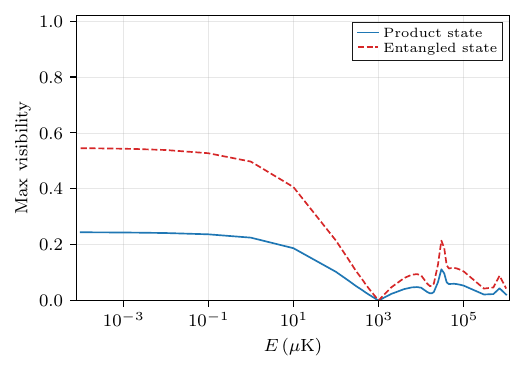}
        \put(0,63){(c)}
    \end{overpic}
    \caption{
        Energy dependence of the maximum visibility for alternative initial channel pairs in product and entangled initial states.
        (a) ${}^6\mathrm{Li}$+${}^6\mathrm{Li}$ collisions prepared in channel pair $(1,2)$
        at $B=3~\mathrm{G}$. This channel exhibits strong resonance structure at low field
        ($0.1~\mathrm{G}$) in the ultracold regime ($E<1~\mu\mathrm{K}$); the higher field
        is chosen to avoid unwanted resonances.
        (b) ${}^7\mathrm{Li}$+${}^7\mathrm{Li}$ collisions prepared in channel pair $(3,8)$
        at $B=0.1~\mathrm{G}$.
        (c) ${}^6\mathrm{Li}$+${}^7\mathrm{Li}$ collisions for channel pair $(2,6)$--$(1,7)$  at $B=0.1~\mathrm{G}$. See Fig.~\ref{fig:channels} for channel definitions.    }
    \label{fig:different_initial_all}
\end{figure*}

To demonstrate that the coherent-control mechanism is not limited to the specific entrance channels used in the main text, we examined alternative choices for the entrance channels. Figure~\ref{fig:different_initial_all} displays the maximum visibility for representative channel selections in $^{6}$Li+$^{6}$Li, $^{7}$Li+$^{7}$Li, and $^{6}$Li+$^{7}$Li. Although the absolute visibility is channel-dependent—with particularly large low-energy visibility for certain $^{6}$Li+$^{7}$Li combinations and the $^{7}$Li+$^{7}$Li pair $(3,8)$—the underlying mechanism of identical-particle symmetry is general and applies to arbitrary initial channel combinations.

\subsection{Shape resonances in Li+Li collisions beyond the ultracold regime}
\label{sec:app_shape_resonance}

\begin{figure*}[tbp]
    \centering
    \includegraphics[width=0.67\columnwidth]{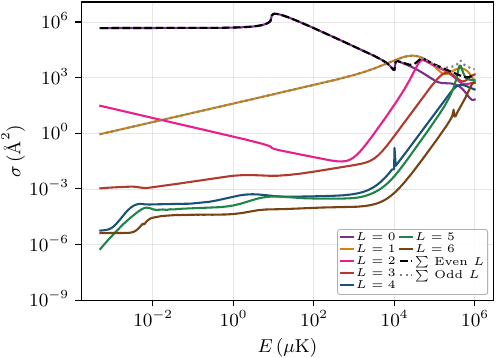}
    \includegraphics[width=0.67\columnwidth]{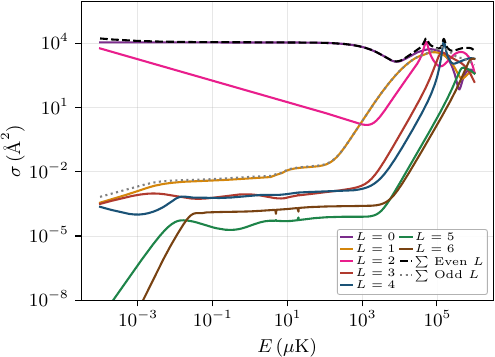}
    \includegraphics[width=0.67\columnwidth]{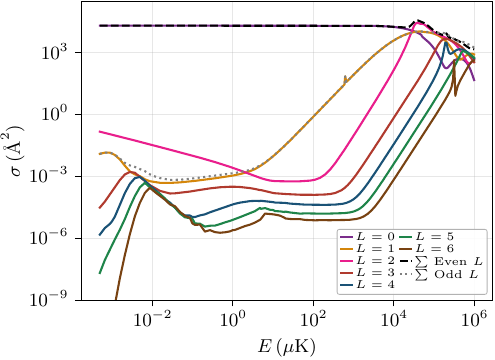}
    \caption{
Total cross sections $\sigma$ for each partial wave $L$, together with the total cross sections summed over the even- and odd-$L$ sectors, at various collision energies $E$ for:
        (a) ${}^6\mathrm{Li}$+${}^6\mathrm{Li}$, 
        (b) ${}^7\mathrm{Li}$+${}^7\mathrm{Li}$,
        (c) ${}^6\mathrm{Li}$+${}^7\mathrm{Li}$. 
    }
    \label{fig:partial_all}
\end{figure*}

In Li+Li collisions, shape resonances arise when a quasi-bound state supported by the centrifugal barrier couples to the scattering continuum within the same partial wave. For instance, the $p$-wave ($L=1$) shape resonance occurs at $\sim 10\,\mathrm{mK}$. Within the coherent-control protocol, these resonances manifest as enhanced maximum visibility and rapid variations in scattering phase near the resonance energy [see Fig.~\ref{fig:visibility_and_partials} of the main text and Fig.~\ref{fig:partial_all}(a) for $^6$Li+$^6$Li].

The corresponding results for $^{7}$Li+$^{7}$Li and $^{6}$Li+$^{7}$Li collisions are shown in Fig.~\ref{fig:partial_all}(b) and (c), respectively. The shape resonances are clearly visible in the total cross sections $\sigma$ for each partial wave $L$, and their locations are consistent with the energy dependence of the maximum visibility shown in Fig.~\ref{fig:visibility_and_partials}(a) in the main text.

\subsection{Effects of magnetic field on coherent control in Li+Li collisions}
\label{sec:app_feshbach}

In coherent control, the magnetic field $B$ can serve as an external tuning parameter that can modify both the magnitudes and phases of the S-matrix elements, thereby affecting the control of the total scattering cross section for a prepared entrance superposition. Previous studies indicate that isolated resonances can enhance control, whereas overlapping resonances often suppress it~\cite{Devolder2023TheJournalofPhysicalChemistryLetters}.

Our coupled-channel calculations for Li+Li show that magnetic Feshbach resonances can significantly affect the achievable visibility at different $E$. A systematic characterization of the full $(B,E)$ dependence is beyond the scope of the present work. Unless stated otherwise, we focus on the near-zero-field regime ($B=0.1~\mathrm{G}$) or the regime far from Feshbach resonances.

\section{General introduction to coherent control of collisions}
\label{sec:app_general_intro}
Coherent control of molecular collisions relies on the preparation of initial states as coherent superpositions of internal molecular degrees of freedom~\cite{Book_Brumer}. In general, such a state can be expressed as
\begin{equation}
\ket{\Psi_{\mathrm{ini}}} = \sum_{i_A=1}^{N_{i,A}} \sum_{i_B=1}^{N_{i,B}} a_{i_A,i_B} \ket{i_A, i_B},
\label{eq:gen_sup}
\end{equation}
where $a_{i_A,i_B} \in \mathbb{C}$ are complex probability amplitudes satisfying the normalization condition $\sum_{i_A,i_B} |a_{i_A,i_B}|^2 = 1$. The internal states $\ket{i_A}$ and $\ket{i_B}$ may correspond to vibrational, rotational, spin, or other relevant degrees of freedom, depending on the physical system under consideration. This expression represents the most general pure state in the product Hilbert space $\mathcal{H}_{\mathrm{int},A} \otimes \mathcal{H}_{\mathrm{int},B}$, and it may exhibit quantum entanglement between the two subsystems. 

Such coherent superpositions give rise to multiple indistinguishable scattering pathways whose associated amplitudes interfere. The resulting interference pattern depends sensitively on the relative phases of the coefficients $a_{i_A,i_B}$, which are determined by the state preparation. Consequently, controlled manipulation of these relative phases enables the modulation of interference effects, providing a mechanism to tune and control observable quantities in the scattering process. It is important to emphasize that the occurrence of interference effects is constrained by the symmetry properties of the collision process. In particular, interference between different pathways is only possible when specific selection rules are satisfied. The following conditions have been identified as necessary for interference between the two states \cite{Book_Brumer}: (a) they must be energetically degenerate, (b) they must satisfy the same total momentum conservation conditions, and (c) they must have the same projection of the total internal angular momentum onto the quantization axis.

As a representative example, consider an initial state prepared as a coherent superposition of two product states,

\begin{equation}
\ket{\Psi_{\mathrm{ini}}} = \cos \theta   \ket{0}_A \ket{1}_B + \sin \theta  e^{i\beta} \ket{1}_A \ket{0}_B .
\end{equation}
The state-to-state integral cross section from this superposition to a given final state $\ket{f}$ can be written as
\begin{equation}
\begin{split}
\sigma_{\mathrm{sup} \to f} = & \cos^2 \theta \ \sigma_{01 \to f} + \sin^2 \theta \ \sigma_{10 \to f} \\
&+ 2 \cos \theta \sin \theta  \left| \sigma_{\mathrm{int},f} \right| \cos(\beta - \Phi_f),
\end{split}
\end{equation}
where the interference contribution is defined as
\begin{equation}
\sigma_{\mathrm{int},f} = \frac{\pi}{k^2} \sum_{\ell,m} \sum_{\ell',m'}
T_{0,1,\ell,m \to f,\ell',m'}
T^{*}_{1,0,\ell,m \to f,\ell',m'}
\end{equation}
and $\Phi_f = \arg(\sigma_{\mathrm{int},f})$ denotes its phase.

The cross section varies sinusoidally with the relative phase $\beta$, providing the basis for coherent control. The degree of control is determined by the relative magnitude of the interference term compared to the incoherent contributions (the first two terms). To quantify this effect, one introduces the visibility,
\begin{equation}
\begin{split}
V = & \frac{\sigma_{\max} - \sigma_{\min}}{\sigma_{\max} + \sigma_{\min}} \\
= & \frac{2 \cos \theta \sin \theta  |\sigma_{\mathrm{int},f}|}{\cos^2 \theta  \sigma_{01 \to f} + \sin^2 \theta \sigma_{10 \to f}} .
\label{eq:visibility_appendix}
\end{split}
\end{equation}

It has been shown that the maximum achievable visibility is given by the ratio~\cite{Devolder2023}
\begin{equation}
R_c = \frac{|\sigma_{\mathrm{int},f}|}{\sqrt{\sigma_{01 \to f}  \sigma_{10 \to f}}},
\end{equation}
referred to as the control index. By construction, $0 \leq R_c \leq 1$. The upper bound $R_c = 1$ is reached when the two pathways are fully indistinguishable, corresponding to equal incoherent contributions. This condition can be expressed as
$P(\theta_{\mathrm{max}}) = \frac{\left| \cos^2 \theta_{\mathrm{max}} \ \sigma_{01 \to f} - \sin^2 \theta_{\mathrm{max}} \ \sigma_{10 \to f} \right|}{\cos^2 \theta_{\mathrm{max}} \ \sigma_{01 \to f} + \sin^2 \theta_{\mathrm{max}} \ \sigma_{10 \to f}} = 0$.

Achieving large values of $R_c$ requires the interference term to approach the geometric mean of the individual cross sections. This condition is closely related to the structure of the control landscape at the level of individual partial-wave transitions. When the interference contributions from all initial–final partial-wave channels add constructively (i.e., are phase-aligned), the control index approaches unity. Conversely, when these contributions are out of phase, destructive competition between channels suppresses the overall interference, leading to reduced controllability and smaller values of $R_c$. 
\section{General coherent-control scattering-matrix formalism}
\label{sec:app_CCS}
Following Refs.~\cite{Frishman1999,Devolder2024TheJournalofChemicalPhysics}, first consider two molecules prepared in a well-defined initial internal state $\ket{\Psi_{\mathrm{ini}}}=\ket{i_A,i_B}$, where $\ket{i_A}$ and $\ket{i_B}$ denote the initial internal quantum states of molecules A and B, respectively. The total cross section is obtained from the corresponding $T$-matrix elements, $T_{i_A,i_B,\ell,m\rightarrow f_A,f_B,\ell',m'}$, as
\begin{equation}
    \sigma =\frac{\pi}{k^2}\sum_{f_A,f_B}\sum_{\ell,m_\ell}\sum_{\ell',m'_\ell}|T_{i_A,i_B,\ell,m_\ell\rightarrow f_A,f_B,\ell',m'_\ell}|^2  
\end{equation}
where $\ket{f_A}$ is an internal state of the first molecule and $\ket{f_B}$ is an internal state of the second molecule after the collision.

Now consider the preparation of the two molecules in the general coherent superposition of internal states in Eq.~\eqref{eq:gen_sup}.
For such an initial state, the total cross section is obtained by coherently summing over all contributing scattering amplitudes:
\begin{equation}
    \sigma =\frac{\pi}{k^2}\sum_{f_A,f_B}\sum_{\ell,m_\ell}\sum_{\ell',m'_\ell}\left|\sum_{i_A, i_B} a_{i_A,i_B}T_{i_A,i_B,\ell,m_\ell\rightarrow f_A,f_B,\ell',m'_\ell} \right|^2 
\end{equation}
Expanding the modulus squared yields a quadratic form in the preparation amplitudes:
\begin{equation}
\begin{split}
    \sigma =&\frac{\pi}{k^2} \sum_{i_A,i_B}\sum_{j_A,j_B} a^*_{i_A,i_B}\\ &\Bigg(\sum_{f_A,f_B}\sum_{\ell,m_\ell}\sum_{\ell',m'_\ell}T_{i_A,i_B,\ell,m_\ell\rightarrow f_A,f_B,\ell',m'_\ell} \\ & \times T^*_{j_A,j_B,\ell,m_\ell\rightarrow f_A,f_B,\ell',m'_\ell}\Bigg) \\ &a_{j_A,j_B}
    \end{split}
\end{equation}

We now reinterpret the $T$-matrix as a rectangular matrix whose rows are indexed by the initial internal states $(i_A,i_B)$ and whose columns collect all remaining quantum numbers $(\ell,m_\ell,f_A,f_B,\ell',m'_\ell)$. With this identification, the quantity inside parentheses corresponds to the matrix product $\boldsymbol{T}\boldsymbol{T}^{\dagger}$. This defines the Coherent Control Scattering (CCS) matrix, $ \mathbf{C} = \mathbf{T}\mathbf{T}^\dagger$, which acts on the space of initial internal states and has dimension $(N_{i_A}N_{i_B})\times(N_{i_A}N_{i_B})$. Its matrix elements are explicitly given by \cite{Devolder2024TheJournalofChemicalPhysics}:
\begin{equation}
\begin{split}   
\mathbf{C}_{i_A,i_B;j_A,j_B}
= \frac{\pi}{k^2}    \sum_{\ell,m_\ell}\sum_{f_A,f_B}\sum_{\ell',m'_\ell} &T_{i_A,i_B,\ell,m_\ell \rightarrow f_A,f_B,\ell',m'_\ell} \\ \times&T^*_{j_A,j_B,\ell,m_\ell \rightarrow f_A,f_B,\ell',m'_\ell}
\end{split}
\end{equation}
This formulation makes explicit that the control of the cross section through coherent superpositions is governed by the Hermitian, positive semidefinite CCS matrix $\boldsymbol{C}$:
\begin{equation}
\sigma = \boldsymbol{a}^\dagger \boldsymbol{C} \boldsymbol{a}
\end{equation}
Here, $\boldsymbol{a}$ is the coefficient vector.


\section{Coherent control of identical-particle collisions in the unsymmetrized basis}
\label{sec:app_unsymmetrized_basis}
In the main text, we developed a formalism describing coherent control of collisions between identical particles using a basis adapted to exchange symmetry. This symmetrized representation naturally separates the dynamics into distinct symmetry sectors and provides a convenient framework to construct the coherent control scattering (CCS) matrix governing the total cross section. In this basis, the CCS matrix takes the diagonal form:
\begin{equation}
\boldsymbol{C}=\begin{pmatrix}
		\sigma_{0,0} & 0 & 0 & 0\\
		0 & \sigma^s_{0,1 } & 0 & 0 \\
		0 & 0 & \sigma^a_{0,1} & 0 \\
		0 & 0 & 0 & \sigma_{1,1}
	\end{pmatrix}
\end{equation}
The CCS matrix can be decomposed into contributions from even- and odd-parity components,
\begin{equation}
\boldsymbol{C}=\begin{pmatrix}
		\sigma_{0,0} & 0 & 0 & 0\\
		0 & \sigma^s_{0,1 } & 0 & 0 \\
		0 & 0 & 0 & 0 \\
		0 & 0 & 0 & \sigma_{1,1}
	\end{pmatrix} +
    \begin{pmatrix}
		0 & 0 & 0 & 0\\
		0 & 0 & 0 & 0 \\
		0 & 0 & \sigma^a_{0,1 }  & 0 \\
		0 & 0 & 0 & 0
	\end{pmatrix}
\end{equation}
As shown in the main text, in the symmetrized basis, the CCS matrix is diagonal, reflecting the fact that states belonging to different exchange-symmetry sectors do not interfere. Consequently, no coherent interference terms arise between distinct symmetry-adapted components of the initial state. While this structure simplifies the analysis of identical-particle scattering, it obscures the interpretation of coherent control in terms of interference and complicates a direct comparison with the case of distinguishable particles.

To facilitate this comparison, we transform back to the unsymmetrized basis using the unitary transformation:
\begin{equation}
\boldsymbol{U}= \begin{pmatrix}
    1 & 0 & 0 & 0 \\
    0 & \frac{1}{\sqrt{2}} & \frac{1}{\sqrt{2}} & 0 \\
    0 & \frac{1}{\sqrt{2}} & -\frac{1}{\sqrt{2}} & 0 \\
    0 & 0 & 0 & 1
\end{pmatrix}
\end{equation}

The state amplitudes and the CCS matrix transform according to
\begin{equation}
\boldsymbol{a}'=\boldsymbol{U}\boldsymbol{a}
\end{equation}
\begin{equation}
\boldsymbol{C}'=\boldsymbol{U}\boldsymbol{C}\boldsymbol{U}^{\dagger}
\end{equation}
Applying this transformation to the even- and odd-parity contributions separately yields
\begin{equation}
\boldsymbol{C}'=\begin{pmatrix}
		\sigma_{0,0} & 0 & 0 & 0\\
		0 & \frac{\sigma^s_{0,1 }}{2} & \frac{\sigma^s_{0,1 }}{2} & 0 \\
		0 & \frac{\sigma^s_{0,1 }}{2} & \frac{\sigma^s_{0,1 }}{2} & 0 \\
		0 & 0 & 0 & \sigma_{1,1}
	\end{pmatrix} +
    \begin{pmatrix}
		0 & 0 & 0 & 0\\
		0 &\frac{\sigma^a_{0,1 }}{2}& -\frac{\sigma^a_{0,1 }}{2}  & 0 \\
		0 & -\frac{\sigma^a_{0,1 }}{2}  & \frac{\sigma^a_{0,1 }}{2} & 0 \\
		0 & 0 & 0 & 0
	\end{pmatrix}
\end{equation}
Each contribution individually satisfies the Cauchy–Schwarz equality, $|C'_{23}|=\sqrt{C'_{22}C'_{33}}$, indicating maximum interference within each symmetry sector. The two contributions correspond to distinct relative phase shifts between the interfering pathways: one with relative phase 0 ($e^{i0}=+1$) and the other with relative phase $\pi$ ($e^{i\pi}=-1$), as shown by the sign of the interference term $C'_{23}$. Moreover, for both contributions individually one finds $C'_{22}=C'_{33}$, reflecting that the two paths are indistinguishable at the level of the transformed basis. Therefore, control is not limited by partial distinguishability between pathways.

Summing the even- and odd-parity contributions yields
\begin{equation}
\boldsymbol{C}'=\begin{pmatrix}
		\sigma_{0,0} & 0 & 0 & 0\\
		0 & \frac{\sigma^s_{0,1}+\sigma^a_{0,1 }}{2} & \frac{\sigma^s_{0,1}-\sigma^a_{0,1}}{2} & 0 \\
		0 & \frac{\sigma^s_{0,1}-\sigma^a_{0,1}}{2} & \frac{\sigma^s_{0,1}+\sigma^a_{0,1 }}{2} & 0 \\
		0 & 0 & 0 & \sigma_{1,1}
	\end{pmatrix}
\end{equation}
In that case, the competition between the two contributions from the two symmetry sectors reduces the controllability. The Cauchy--Schwarz equality is no longer satisfied: $|C'_{23}|=\frac{|\sigma^s_{0,1}-\sigma^a_{0,1}|}{2}< \sqrt{C'_{22}C'_{33}}=\frac{\sigma^s_{0,1}+\sigma^a_{0,1 }}{2}$. Complete coherent control is recovered in regimes where either the even- or odd-parity channel is dominant, such as in the ultracold limit where a single partial-wave contribution typically prevails.

\section{Initial-state preparation: product versus entangled superpositions}
\label{sec:app_initial_states}

In addition to the product-state superposition

\begin{equation}
    \label{eq:sup_prod}
    \begin{split}
    \ket{\Psi_{\mathrm{ini}}} = &\left(\cos \theta\ket{0}+\sin\theta\ket{1}\right)_{\!A} \\
    &\otimes\left(\cos\theta\ket{0}+e^{i\beta}\sin\theta\ket{1}\right)_{\!B}
    \end{split}
    \end{equation}
analyzed in the main text, one can prepare entangled states that populate exclusively the symmetric and antisymmetric combinations in the symmetrized basis:

\begin{equation}
\label{eq:ini_ent}
	\ket{\Psi_{\mathrm{ini}}}=\frac{\cos\theta + \sin\theta  e^{i\beta}}{\sqrt{2}} \ket{0,1}_s+\frac{\cos\theta - \sin\theta  e^{i\beta}}{\sqrt{2}}\ket{0,1}_a
\end{equation}
where the specific states $0$ and $1$ used in the paper are defined in Table~\ref{tab:twolevel_map}.

These entangled initial states remove the satellite terms from the total scattering cross sections~\cite{Devolder2025} and are expected to enhance the visibility across different collision energies. However, the experimental preparation of entangled initial states is significantly more challenging than that of the product states defined in Eq.~\eqref{eq:sup_prod}. 

For the case of non-identical particles, the total cross section can be expressed as
\begin{equation}
\begin{split}
\sigma^{\mathrm{non-ident}}(\theta, \beta)=&\cos^2\theta \sigma_{0,1}  +\sin^2\theta \sigma_{1,0}\\+& 2\sin\theta \cos\theta \sum_f |\sigma_{\mathrm{int},f}| \cos(\Phi_f+\beta),
\end{split}
\label{eq:cont_total_non_ident_ent}
\end{equation}
For identical particles, the total cross section becomes
\begin{equation}
\sigma^{\mathrm{ident}}(\theta,\beta)=\frac{|\cos \theta+\sin\theta e^{i\beta}|^2}{2} \sigma^s_{0,1}+\frac{|\cos \theta-\sin\theta e^{i\beta}|^2}{2} \sigma^a_{0,1} 
\label{eq:coherent_control_sigma_ent}
\end{equation}
The visibility of the control over the total cross section, evaluated at $\theta=\pi/4$, becomes
\begin{equation}
\begin{split}
V(\pi/4)= &\frac{\sigma_{\mathrm{max}}(\theta,\beta_{\mathrm{max}})-\sigma_{\mathrm{min}}(\theta,\beta_{\mathrm{min}})}{\sigma_{\mathrm{max}}(\theta,\beta_{\mathrm{max}})+\sigma_{\mathrm{min}}(\theta,\beta_{\mathrm{min}})} \\
=&\frac{\left|\frac{\sigma_{01}^s}{\sigma_{01}^a}-1\right|}{\frac{\sigma_{01}^s}{\sigma_{01}^a}+1}
\end{split}
\label{eq:Svisibility_ent}
\end{equation}
For entangled superposition states, the visibility depends solely on the relative weight of the symmetric and antisymmetric contributions. Importantly, it is no longer constrained by the satellite terms. In the ultracold regime, one of these contributions typically dominates over the other, enabling complete control of the total cross section. Thus, for entangled superpositions, full control of the total cross section at ultracold temperatures can be achieved for both identical fermions and identical bosons. This contrasts with the product-state case, where complete control is not attainable for identical bosons due to the presence of satellite terms.

\begin{figure*}[tbp]
\centering
\includegraphics[width=\textwidth]{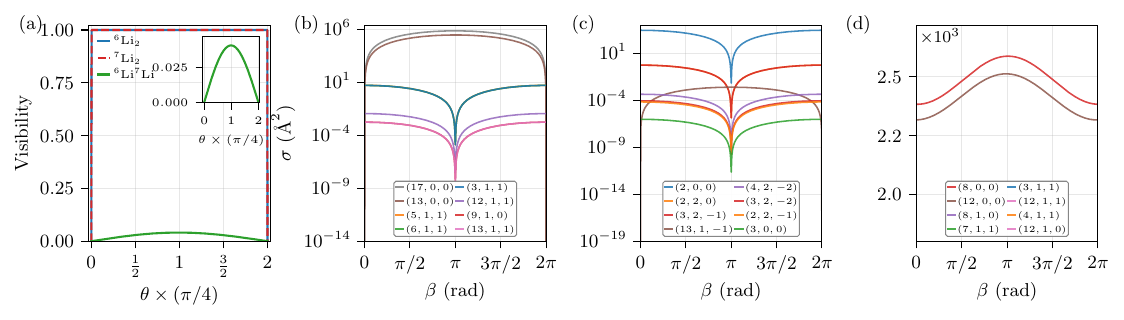}
\caption{(a) Visibility $V$ of the total cross-section control versus $\theta$ for $^{6}$Li+$^{6}$Li, $^{7}$Li+$^{7}$Li, and $^{6}$Li+$^{7}$Li with entangled initial states; the inset zooms in for $^{6}$Li+$^{7}$Li.
(b)--(d) Dominant partial-wave-resolved state-to-state cross sections for (b) $^{6}$Li+$^{6}$Li, (c) $^{7}$Li+$^{7}$Li, and (d) $^{6}$Li+$^{7}$Li with entangled initial states.
Labels denote exit-channel quantum numbers $(\alpha_{\rm out},L_{\rm out},m_{L,\rm out})$; colors indicate partial waves and opacity indicates hyperfine exit channels.}
\label{fig:totaltheta_entangled}
\end{figure*}

\begin{figure*}[tbp]
\centering
\includegraphics[width=\textwidth]{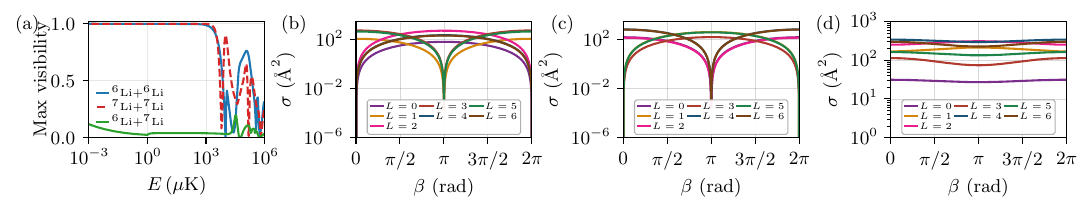}
\caption{(a) Maximum visibility $V$ of the total cross-section control versus collision energy $E$ for $^{6}$Li+$^{6}$Li, $^{7}$Li+$^{7}$Li, and $^{6}$Li+$^{7}$Li with entangled initial states.
(b)--(d) Total cross sections at different partial waves at $E=1\,\mathrm{K}$ with entangled initial states for
(b) $^{6}$Li+$^{6}$Li, (c) $^{7}$Li+$^{7}$Li, and (d) $^{6}$Li+$^{7}$Li.}
\label{fig:supp_entangled}
\end{figure*}

Figures~\ref{fig:totaltheta_entangled} and \ref{fig:supp_entangled} present the entangled-initial-state analogs of the results shown in the main text.

Figure~\ref{fig:totaltheta_entangled} shows the visibility for control of the total cross section versus mixing angle $\theta$ for $^{6}$Li+$^{6}$Li, $^{7}$Li+$^{7}$Li, and $^{6}$Li+$^{7}$Li at $E=20~\mu\mathrm{K}$. For $^{6}$Li+$^{6}$Li collisions, the visibility remains 1 across all control angles $\theta$. In contrast to the limited maximum visibility of $\sim 0.7$ with the product initial state, $^{7}$Li+$^{7}$Li shows complete control with $V=1$ for all $\theta$ at $E=20\,\mu\mathrm{K}$, since the removal of the satellite terms boosts the visibility. For $^{6}$Li+$^{7}$Li, although we see a slight increase in maximum visibility from 0.02 to $\sim 0.05$, it is still much smaller than the corresponding values for the homonuclear cases. This is because the lack of phase-locking in $^{6}$Li+$^{7}$Li occurs even within each symmetry sector.

The partial scattering cross sections with entangled initial states at $E=20\,\mu\mathrm{K}$ are shown in Fig.~\ref{fig:totaltheta_entangled}(b)--(d). For identical fermions, the figure resembles the one in the main text, since the satellite terms are naturally suppressed at ultracold temperatures. For identical bosons, the enhancement of the visibility is shown clearly in panel (c), where only the symmetric components contribute to the total scattering cross section with the satellite term suppressed, yielding $V=1$. 

\begin{figure}[tbp]
    \centering
    \includegraphics[width=\columnwidth]{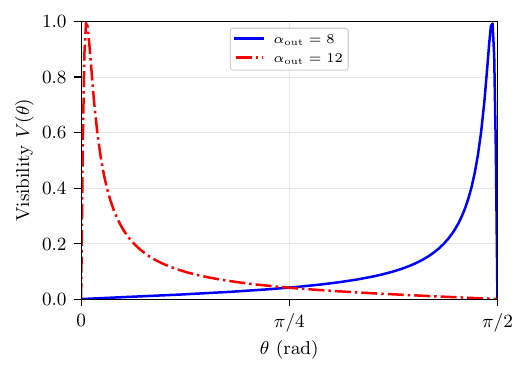}
\caption{
Visibility $V(\theta)$ for the dominant state-to-state channels in the $^{6}$Li+$^{7}$Li collision at $E=2\times10^{-5}\,$K as the mixing angle $\theta$ (which sets the relative populations of the two initial components $\ket{01}$ and $\ket{10}$) is varied. Because the single-path cross sections are asymmetric, the two channels require opposite population imbalances to balance the interfering amplitudes and thus enhance indistinguishability: for the final state $(8,0,0)$ one has $\sigma_{01\rightarrow 8}>\sigma_{10\rightarrow 8}$, so increased population in $\ket{10}$ (blue curve) is needed to maximize interference, whereas for $(12,0,0)$ the asymmetry is reversed, $\sigma_{10\rightarrow 12}>\sigma_{01\rightarrow 12}$, and increased population in $\ket{01}$ (red curve) is required. Consequently, the two channels cannot be optimized simultaneously, which limits the visibility in the total cross section; in particular, near $\theta=\pi/4$ (equal pathway weights) both channels exhibit weak control due to significant pathway distinguishability. 
}

    \label{fig:6Li7Li_distinguishability}
\end{figure}

Panel (d) clearly shows weak control for both populated final states, $(8,0,0)$ and $(12,0,0)$. This behavior originates from the partial distinguishability of the two interfering pathways and from the asymmetry of the corresponding transition cross sections. For the final state $(8,0,0)$, the cross section from $\ket{01}$, $\sigma_{01 \rightarrow 8}$, is larger than that from $\ket{10}$, $\sigma_{10 \rightarrow 8}$. To enhance interference and thus improve control, the amplitudes of the two pathways must be balanced. This would require increasing the population in the state $\ket{10}$, as illustrated by the blue curve in Fig.~\ref{fig:6Li7Li_distinguishability}. In contrast, for the final state $(12,0,0)$, the asymmetry is reversed: $\sigma_{10 \rightarrow 12}$ exceeds $\sigma_{01 \rightarrow 12}$. Achieving balanced pathway contributions in this case would therefore require increasing the population in $\ket{01}$, as shown by the red curve in Fig.~\ref{fig:6Li7Li_distinguishability}. Consequently, when considering the total cross section, it is impossible to simultaneously render the two pathways indistinguishable for both channels. This intrinsic incompatibility leads to the reduced visibility observed in the total cross section [Fig.~\ref{fig:totaltheta_entangled}(a)]. In particular, at $\theta = \pi/4$, Fig.~\ref{fig:totaltheta_entangled}(d) shows weak control for both final states, reflecting the significant distinguishability of the two interfering pathways.

The energy dependence of the visibility for entangled initial states is shown in Fig.~\ref{fig:supp_entangled}. For $^{6}$Li+$^{6}$Li and $^{7}$Li+$^{7}$Li, the introduction of the entangled initial state suppresses the contributions of the satellite terms at higher collision energies beyond the $s$-wave limit. Thus, the maximum visibilities of the homonuclear cases increase between $\sim1\,\mathrm{mK}$ and $1\,\mathrm{K}$. The $^{7}$Li+$^{7}$Li system then has complete controllability over a wider range of collision energies, similar to $^{6}$Li+$^{6}$Li, until it reaches the $p$-wave centrifugal barrier near $\sim10\,\mathrm{mK}$. For $^{6}$Li+$^{7}$Li, even though the satellite term is suppressed and the overall energy dependence of maximum visibility is enhanced, the maximum visibility is still small across different collision energies because of the lack of phase-locking.

\section{Identical-particle collisions with different single-particle superpositions}
\label{sec:app_different states}
Consider the scenario where the first molecule is prepared in a superposition of states $\ket{0}$ and $\ket{1}$ and the second molecule is prepared in a superposition of states $\ket{1}$ and $\ket{2}$:
\begin{equation}
	\ket{\psi_a}=\cos \theta \ket{0} + \sin \theta  \ket{1}
    \label{eq:Ssup_first}
\end{equation}
\begin{equation}
	\ket{\psi_b}=\cos \theta \ket{1} + \sin \theta e^{i\beta}  \ket{2}
    \label{eq:Ssup_second}
\end{equation}
This kind of superposition was considered in our previous work on coherent control of ultracold collisions between O$_2$ molecules \cite{Devolder2021Phys.Rev.Lett.}. 
The total internal state of the two-molecule system in the symmetrized basis is
\begin{equation}
\begin{split}
\ket{\Psi_{\mathrm{ini}}}= & \frac{\cos^2\theta}{\sqrt{2}}\left(\ket{01}_a+\ket{01}_s \right)+\frac{\cos\theta\sin\theta e^{i\beta}}{\sqrt{2}}\left(\ket{02}_a+\ket{02}_s \right) \\ +& \cos\theta\sin\theta \ket{11} + \frac{\sin^2\theta e^{i\beta}}{\sqrt{2}}\left(\ket{12}_a+\ket{12}_s \right)
\end{split}
\end{equation}
The first notable difference is that the population of each component, given by the squared magnitude of its coefficient, is independent of the phase parameter $\beta$.

Second, interference effects can arise in the symmetrized basis. For instance, if the states are chosen such that $\ket{02}$ and $\ket{11}$ are degenerate in energy and share the same total projection of internal angular momentum (as in Ref.~\cite{Devolder2021Phys.Rev.Lett.}), then the states $\ket{11}$ and $\ket{02}_s$ can coherently interfere. Both states are symmetric under particle exchange and therefore scatter through the same set of partial waves (even for bosons and odd for fermions). In this case, the CCS matrix is no longer diagonal.

 The total cross section in this case becomes
 \begin{equation}
\begin{split}
\sigma(\theta, \beta)=&\frac{\cos^4\theta}{2} (\sigma_{01,a}+\sigma_{01,s})+\frac{\cos^2\theta\sin^2\theta}{2} (\sigma_{02,a}+\sigma_{02,s})\\+ &\cos^2\theta \sin^2\theta \sigma_{1,1}+\frac{\sin^4\theta}{2} (\sigma_{12,a}+\sigma_{12,s})\\+& \frac{2}{\sqrt{2}}\sin^2\theta \cos^2\theta \sum_f |\sigma_{\mathrm{int},f}| \cos(\Phi_f+\beta),
\end{split}
\label{eq:Scont_total_non_ident}
\end{equation}
where $\sigma_{\mathrm{int},f}=\boldsymbol{C}^{\mathrm{sup} \rightarrow f}_{02s,11}$.

Importantly, the interference term between $\ket{11}$ and $\ket{02}_s$ is not constrained by exchange symmetry to be identical for all final states. Consequently, one recovers a situation analogous to the non-identical-particle case, where the control parameter depends on the specific final state. Upon summation over final states, this leads to partial scrambling of the interference effects and a corresponding loss of global control.

The synchronization of the control across all final states, the main result of this work, requires one to use the initial superposition composed of the same internal eigenstates (Eqs.~\eqref{eq:sup_first} and \eqref{eq:sup_second} of the main text).


\section{Behavior of state-to-state cross sections at higher collision energies}
\label{sec:app_energy_sync}

\begin{figure*}[t]
\centering
\includegraphics[width=\textwidth]{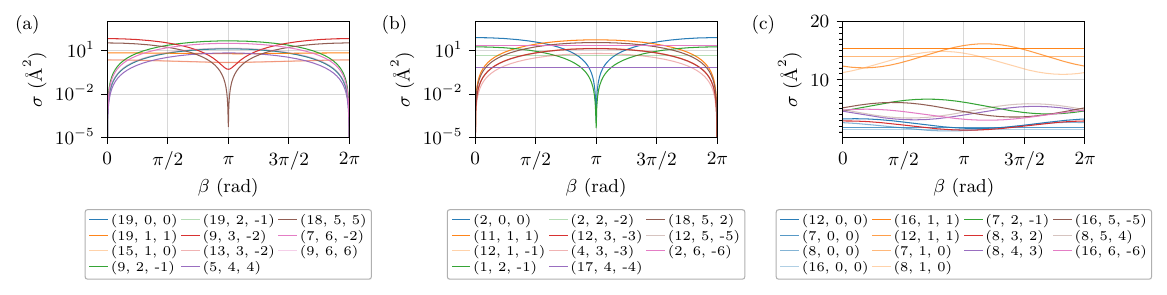}
\caption{
Partial-wave-resolved state-to-state cross sections at $E=1\,\mathrm{K}$ with product initial states for (a) $^{6}$Li+$^{6}$Li, (b) $^{7}$Li+$^{7}$Li, and (c) $^{6}$Li+$^{7}$Li.
Labels denote exit-channel quantum numbers $(\alpha_{\rm out},L_{\rm out},m_{L,\rm out})$; colors indicate partial waves and opacity indicates hyperfine exit channels.
}
\label{fig:partialsigma_energy}
\end{figure*}

In this section, we analyze the mechanism responsible for the reduced control visibility of the total scattering cross sections shown in Fig.~\ref{fig:visibility_and_partials} of the main text. We show that, for identical particles, the loss of visibility arises primarily from the incoherent summation of contributions from different parity sectors. Within each symmetry sector, however, the control phases of the hyperfine state-to-state cross sections remain locked. In contrast, for distinguishable particles, phase scrambling occurs not only between different partial-wave parity sectors, but also within each sector, due to the absence of exchange symmetry.

As shown in Fig.~\ref{fig:partialsigma_energy}, at $1\,\mathrm{K}$ many state-to-state cross sections contribute to the total scattering cross section with comparable magnitudes, making coherent control more difficult. For the homonuclear systems, $^{6}$Li+$^{6}$Li and $^{7}$Li+$^{7}$Li, exchange symmetry locks the beta-dependent modulation of the state-to-state cross sections within a given parity sector. As a result, all contributions belonging to the same parity sector share the same control phase, independent of the specific final hyperfine state or partial wave. This phase-locking is evident in panels (a) and (b). The total visibility is reduced only when even- and odd-parity contributions, which have different phases, are summed together.

The heteronuclear $^{6}$Li+$^{7}$Li system behaves qualitatively differently. Because the particles are distinguishable, both even and odd partial waves are allowed for the same hyperfine entrance channel, and exchange symmetry no longer imposes phase constraints. Consequently, the relative phases of different state-to-state cross sections depend on both the final hyperfine channel and the partial wave $(L,m_L)$, leading to both internal-state and partial-wave scrambling. This is clearly seen in panel (c): within a fixed partial-wave sector, such as $L=0$ or $L=1$, different final internal states exhibit different control phases. Conversely, for a fixed final internal state, such as a state with $\alpha_{\rm out}=8$, the control phase varies across different partial waves, including $L=0,1,3,4,$ and $5$. This behavior is absent in the homonuclear cases, where all channels within the same parity sector remain phase-locked.

\input{new.bbl}

\end{document}

%% file: new.bbl
%